\documentclass{article}
\usepackage{microtype}
\usepackage{graphicx}
\usepackage{subfigure}
\usepackage{booktabs}
\usepackage{amsmath}
\usepackage{amssymb}
\usepackage{tikz}
\usetikzlibrary{arrows.meta,calc,fit,positioning,shapes.geometric}
\usepackage{hyperref}

\usepackage[accepted]{mlsys2025}
\makeatletter
\renewcommand{\Notice@String}{}
\makeatother

\mlsystitlerunning{\footnotesize GRACE: Generative Recommender Acceleration Engine for Real-Time Ads Retrieval}

\newcommand{\grace}{\textsc{GRACE}}
\newcommand{\sid}{SID}
\newcommand{\sids}{SIDs}
\newcommand{\cd}{CD}

\newcommand{\gtm}{GTM}
\newcommand{\Description}[1]{}

\newcommand{\widetable}{\footnotesize\setlength{\tabcolsep}{2pt}\renewcommand{\arraystretch}{0.92}}
\tikzset{
  gracebox/.style={draw, rounded corners=3pt, line width=0.8pt, align=center,
    minimum height=0.44in, inner xsep=7pt, inner ysep=5pt, font=\small},
  graceflow/.style={-{Latex[length=3mm,width=2mm]}, line width=0.95pt},
  gracegroup/.style={draw, rounded corners=3pt, dashed, line width=0.75pt,
    inner sep=7pt}
}

\begin{document}

\twocolumn[
\mlsystitle{GRACE: Generative Recommender Acceleration Engine for Real-Time Ads Retrieval}

\begin{mlsysauthorlist}
\mlsysauthor{Zhou Fang}{meta}
\mlsysauthor{Yuhang Huang}{meta}
\mlsysauthor{Ang Zhang}{meta}
\mlsysauthor{Yihan He}{meta}
\mlsysauthor{Ruichao Xiao}{meta}
\mlsysauthor{Chao Li}{meta}
\mlsysauthor{Yavuz Yetim}{meta}
\mlsysauthor{Sibyl Yang}{meta}
\mlsysauthor{Xiaohan Wei}{meta}
\mlsysauthor{Fei Tian}{meta}
\mlsysauthor{Liang Wang}{meta}
\mlsysauthor{Chonglin Sun}{meta}
\mlsysauthor{Liyuan Li}{meta}
\mlsysauthor{Nathan Yan}{meta}
\mlsysauthor{Gaoxiang Liu}{meta}
\end{mlsysauthorlist}

\mlsysaffiliation{meta}{Meta Platforms}
\mlsyscorrespondingauthor{Zhou Fang}{zhoufang@meta.com}

\mlsyskeywords{generative retrieval, ads retrieval, Transformer, beam search,
constrained decoding, target matching, GPU inference, attention kernels,
KV cache}

\vskip 0.3in

\begin{abstract}
Productionizing generative recommenders for high-volume, real-time ads
retrieval creates two serving challenges: \textbf{eligibility}, ensuring that each generated
ad is eligible for the request under the advertiser's audience targeting rules,
and \textbf{compute}, which requires meeting strict latency and GPU cost
requirements while remaining capable of generating thousands of ads per request
with wide-beam decoding. This paper presents \textsc{GRACE}, a serving system
for ads generative retrieval that addresses both challenges. For eligibility,
\textsc{GRACE} introduces Generative Target Matching (\gtm), which extends
catalog-valid constrained decoding with personalized filtering over Semantic ID (\sid)
prefixes using bitmask and Bloom filter matchers derived from targeting rules.
SID-level \gtm{} improves final ad-level target matching pass rate from 23.55\% to 40.42\% over
constrained decoding alone. For compute-cost and latency, \textsc{GRACE} targets
encoder-decoder Transformers, which are more lightweight than LLMs. It
redesigns the decoder around the wide-beam, short-sequence regime, covering
attention kernels, KV cache, and beam search optimizations. On NVIDIA GH200,
compared with the faster of FlashAttention-2 and FlashAttention-3 baselines,
\textsc{GRACE} improves cross-attention latency by 68.0$\times$ and self-attention latency by
23.4--25.8$\times$ across decode steps. Together, these changes reduce decoder latency
by 11.1$\times$, keeping ads generative retrieval within latency and compute requirements.
\end{abstract}
]

\printAffiliationsAndNotice{}

\section{Introduction}

Large-scale ads recommender systems commonly use a cascaded retrieval and
ranking architecture: fast retrieval stages narrow a large ad corpus to a
smaller candidate set, and more heavyweight ranking models score those
candidates~\cite{youtube,cascade_ranking}. Ads recommendation adds a hard
eligibility requirement. Advertisers can specify targeting rules
over attributes such as geography, demographics, and other request signals to
define the audience they intend to reach. For each user request, ads candidate
generation must enforce these rules so that only ads eligible for that user
enter retrieval. Retrieval then selects a smaller set from this eligible
inventory for downstream ranking. We refer to this eligibility check as Target
Matching.

Generative retrieval changes this flow. Instead of scoring candidates produced
by Target Matching, a Transformer decoder can represent each item as a Semantic
ID (\sid), a short sequence of discrete tokens, and autoregressively emit
\sid{} sequences conditioned on the user~\cite{tiger,plum,onerec}. This creates
a serving mismatch for ads: targeting rules change dynamically and define a
different eligible ad subset for each request, while the generative model learns
a stable \sid{} space and predicts high-probability identifiers directly.
A post-generation filter can expand generated \sids{} to ads and then run Target
Matching, but many generated ads may be discarded before reaching the downstream
ranking stage.

This mismatch makes \textbf{eligibility} the first serving problem for
ads generative retrieval. Catalog-valid constrained decoding does not solve it
by itself. STATIC~\cite{static}, for example, efficiently enforces that each
generated \sid{} belongs to a constrained catalog. That is a validity constraint:
decoding is restricted to SIDs that exist in the catalog, and can support request-independent
business rules such as freshness, locality, category, or inventory constraints.
For ads, generation can also require a personalized eligibility
constraint: the generated \sid{} must represent at least one ad whose targeting
rules match the user. Thus, eligibility must be checked during decoding,
not only after generated SIDs are expanded back to ads.

\textbf{Compute} is the second serving problem. Real-time ads retrieval needs
the capability to generate thousands of ads per request under strict latency and
compute-cost requirements. This motivates our serving design around a lightweight encoder-decoder
Transformer~\cite{transformer,tiger} rather than an LLM.
Its decoder shape differs from the long-sequence regimes targeted by general
attention kernels such as FlashAttention and FlashInfer~\cite{flashattention,fa2,fa3,flashinfer}:
beam search creates a wide effective batch, self-attention sequences are only a
few \sid{} tokens long, and encoder outputs are shared across beams as
cross-attention context. This wide-batch, short-sequence shape leaves general attention
kernels severely underutilized, requiring specialized decoder optimizations.

This paper presents \grace{} (Generative Recommender ACceleration Engine), a
serving system for ads generative retrieval that addresses both problems. For
\textbf{eligibility}, \grace{} introduces \emph{Generative Target Matching}
(\gtm), which preserves catalog-valid decoding while adding a request specific
targeting test at each decode step. It uses bitmask matchers for
low-cardinality attributes such as country, age, and gender, and Bloom filter
matchers for high-cardinality attributes such as location.
Candidate tokens are masked unless they both extend a valid \sid{} prefix
and preserve at least one eligible ad under that prefix,
so beam search advances only through prefixes that remain eligible for
the request. In an evaluation over 30M \sids{}, \sid{}-level \gtm{} improves
final ad-level target matching pass rate from 23.55\% to 40.42\% over
constrained decoding alone.

For \textbf{compute}, \grace{} redesigns the decoder with optimized
cross-attention layout, short-sequence self-attention kernels, paged KV cache
for beam search, \gtm{} kernels, and dynamic per-step beam sizes. On NVIDIA
GH200, compared with the faster of FlashAttention-2 and FlashAttention-3 scaled
dot-product attention (SDPA) baselines, \grace{} improves cross-attention latency by
68.0$\times$ and self-attention latency by 23.4--25.8$\times$ across decode steps. Together,
these optimizations reduce fixed beam size \(M=1024\) decoder latency by
11.1$\times$, from 197.7 ms to 17.8 ms; dynamic per-step beams
further reduce optimized decoder latency to 15.8 ms. With
these changes, \grace{} keeps generative retrieval within the latency and
compute requirements of real-time ads retrieval even with \gtm{} overhead.

This paper makes three contributions:
\begin{enumerate}
  \item It identifies the gap between catalog-valid constrained decoding and
  ads audience target matching, and introduces \gtm{} for personalized
  eligibility testing inside Transformer decoding.
  \item It optimizes the wide-beam, short-sequence attention of
  encoder-decoder generative retrievers with custom kernels that substantially
  outperform state-of-the-art general attention kernels for this workload shape.
  \item It provides an end-to-end serving design for ads generative retrieval,
  including multi-stage pipelined inference and beam search optimizations.
\end{enumerate}

\begin{figure}[t]
  \centering
  \includegraphics[width=\columnwidth]{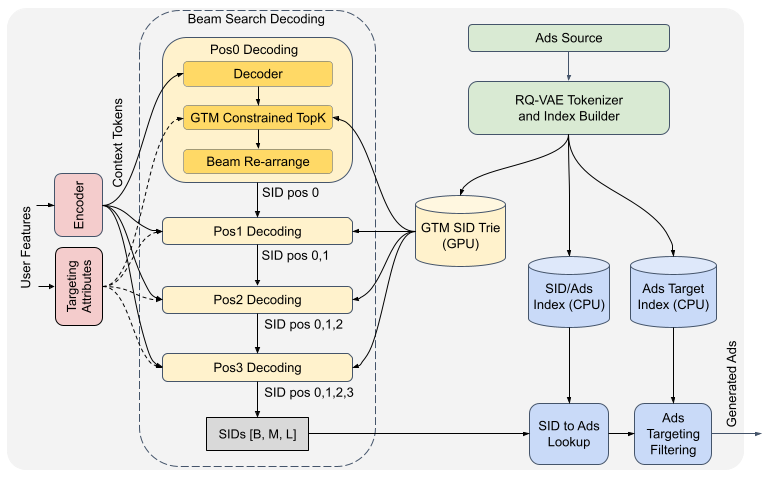}
  \caption{\grace{} serving path. Offline, ads source data is used to asynchronously
  build the GTM SID trie, the SID-to-ad index, and the ad-level target matching index.
  Online serving converts user features into encoder context tokens and runs four-step
  beam search decoding. At each SID position, \gtm{} constrained
  top-\(k\) enforces both valid SID transitions and personalized target
  matching against user-side targeting attributes. Post-processing expands
  generated SIDs to ads and applies exact ad-level target filtering.}
  \label{fig:system-overview}
\end{figure}

\section{RELATED WORK}

\subsection{Ads Recommendation and Audience Targeting}

Conventional ads serving is organized as a cascade. First, Target Matching
enforces advertiser audience rules: for each request, the system checks user and
request attributes against targeting constraints and keeps only ads eligible for
that request. Second, retrieval selects a much smaller candidate set from this
eligible inventory under strict latency constraints.
Third, ranking stages apply richer models to the retrieved
candidates. Andromeda is a representative
large-scale ads retrieval system, combining deep retrieval models with
hierarchical indexing to search over tens of millions of ad candidates
\cite{andromeda}. Work on early-stage ads ranking studies the tradeoff between
efficient pre-ranking and final-stage ranking quality~\cite{early_stage}, while
broader ranking and recommendation models improve feature interaction, user
modeling, and large-scale architecture design
\cite{dlrm,dcnv2,dhen,interformer,hstu,gem}.

\grace{} differs from this conventional flow because a generative retriever
emits ad identifiers directly instead of scoring candidates after Target
Matching. As a result, ads eligibility cannot be enforced only before retrieval;
it must be enforced inside the generative decoding loop.

\subsection{Generative Retrieval}

Generative retrieval replaces nearest-neighbor or candidate-scoring retrieval
with autoregressive generation of identifiers. GENRE generates entity names for
entity retrieval~\cite{genre}, and DSI maps queries directly to document
identifiers with a Transformer index~\cite{dsi}. TIGER adapts this formulation
to recommendation by assigning each item a Semantic ID (\sid) and generating
the next item's \sid{} from user history~\cite{tiger}. Subsequent systems study
SID design, LLM adaptation, and industrial-scale generative
recommendation, including LC-Rec~\cite{lcrec}, PLUM~\cite{plum}, and OneRec~\cite{onerec}.
These works primarily address tokenization, training, alignment, and retrieval quality.
\grace{} is complementary: it studies how to serve \sid{}-based ads retrieval when generated
identifiers must satisfy request-time targeting rules under strict latency and
compute-cost requirements.

\subsection{Constrained Decoding}

Constrained decoding (\cd{}) restricts autoregressive generation at inference
time, with prior work enforcing lexical predicate-logic constraints~\cite{neurologic}
and code-validity constraints such as syntax, typing, and contextual
logic~\cite{synchromesh}. In generative retrieval, the central constraint
is catalog validity: the generated identifier must belong to the allowed item
set. DISC-PPV accelerates prefix verification for constrained generation
\cite{disc_ppv}, and STATIC provides accelerator-friendly constrained decoding
for catalog-valid \sids{}~\cite{static}. \grace{} builds on this direction but
targets a different constraint. Catalog validity is request-independent, while
ads eligibility is personalized: the feasible next token depends on both the
\sid{} prefix and the user's targeting attributes. \gtm{} therefore augments
catalog-valid decoding with request-specific target matching before beam
selection.

\subsection{Attention Kernels}

Efficient attention kernels and LLM serving runtimes optimize Transformer
inference across important production regimes. The
FlashAttention family improves attention through IO-aware tiling,
sequence-parallel work partitioning, and hardware-specific pipeline co-design for
Hopper and Blackwell GPUs~\cite{flashattention,fa2,fa3,fa4}. LLM serving systems
such as vLLM/PagedAttention and FlashInfer focus on high-throughput
autoregressive serving through dynamic batching, KV cache memory management,
customizable attention, and decode-time scheduling~\cite{pagedattention,flashinfer}.
\grace{} instead targets a wide-beam, short-sequence decoder with shared encoder context tokens,
and decode-time \cd+\gtm{} matching. These differences motivate kernel optimizations specialized
for this workload.

\section{GRACE Serving Architecture}

\subsection{Inference Flow}
\label{sec:inference-flow}

\grace{} provides a serving framework for placing generative retrieval within a
multi-stage real-time ads serving stack. Figure~\ref{fig:system-overview}
summarizes the inference flow. First, an encoder converts request-side user
features into context tokens. Second, \grace{} runs an autoregressive
beam search loop over SID positions. At each position, the decoder produces
next-token logits for the current beams; a fused constraint kernel applies both
\cd{} and \gtm{} filtering to remove invalid or ineligible next tokens. For each
surviving next token, the beam score is updated by adding the new token's log-probability to the
accumulated log-probability of the prefix; beam top-$k$ then selects the
highest-scoring prefixes and rearranges beam state for the next decode step.
Third, the completed SIDs are expanded to ads through a \sid{}-to-ad index. Fourth, the
expanded ads pass through the ad-level exact target matcher. The
remaining ads then continue to the downstream retrieval and ranking stack.

The flow uses three physical indexes, built periodically and asynchronously
from the real-time online flow, as shown in Figure~\ref{fig:system-overview}.
\begin{itemize}
  \item \textbf{GTM SID trie.} A trie-based constrained decoding index on GPU,
  following STATIC~\cite{static}, augmented with targeting matchers for SID
  prefixes and full SIDs. Low-cardinality
  targeting attributes, such as country, age, and gender, use bitmask
  matchers; high-cardinality targeting attributes, such as fine-grained
  locations, use Bloom filter matchers. This is the only index read by the \gtm{}
  kernel inside the decoding loop.
  \item \textbf{SID-to-ad index.} A CPU-resident map from completed SIDs to
  ad ids. \cd{} ensures that all generated SIDs are catalog-valid, so each
  generated SID expands to at least one generated ad.
  \item \textbf{Ads targeting index.} A CPU-resident ad-level target
  matcher, which enforces the same bitmask and Bloom filter targeting semantics
  as \gtm{} but at exact ad granularity.
\end{itemize}

This CPU-side post-matching remains necessary because \gtm{} operates at \sid{}
granularity, while each \sid{} is a cluster of ads. An eligible \sid{} only
guarantees that at least one ad in the cluster is eligible for the request; the
other ads expanded from the same \sid{} may still be ineligible and must be
removed by exact ad-level filtering. With \gtm{}, however, this post-matching
stage receives SIDs that have already passed personalized eligibility checks
during decoding, so its pass rate is much higher than with ad-level target matching
alone (Section~\ref{sec:eval-gtm-kway}).

\subsection{Encoder-Decoder Model}
\label{sec:encoder-decoder-model}

\gtm{} works for SID-based generative retrievers, including encoder-decoder and
decoder-only Transformer models, while the compute optimizations in this
paper focus on encoder-decoder models~\cite{transformer}:
an encoder maps request features and histories to cross-attention context tokens,
and a decoder autoregressively generates the SID.

\textbf{Encoder.} \grace{} is agnostic to the encoder choice as long as it
emits context tokens for decoder cross-attention. TIGER demonstrates this
encoder-decoder pattern: its Transformer encoder
encodes a user's historical item sequence, and its decoder generates the next
SID autoregressively~\cite{tiger}. For heterogeneous recommendation features,
InterFormer provides an encoder design that combines non-sequence features,
such as user profile and context, with behavior sequences, such as clicks and
conversions, through feature-interaction and attention layers~\cite{interformer}.
Our evaluation in Section~\ref{sec:evaluation} uses this InterFormer-style encoder, which emits
\(T_{\mathrm{ctx}}=128\) context tokens.

\textbf{Decoder.} In beam search decoding, all beams for the same request share
the encoder outputs, so cross-attention can reuse the same KV across beams.
The decoder maintains separate KV caches for cross-attention and
self-attention. Cross-attention KV is computed once per request from encoder
context tokens and is shared by all beams for that request throughout decoding.
Self-attention KV stores each layer's generated-prefix key/value tensors and is
updated after each autoregressive step as the partial SID grows.
The decoder self-attention sequence is short because it contains only the generated
SID prefix, not a long prompt prefix as in decoder-only LLMs.

In the model studied here, the SID has fixed length \(L_{\mathrm{SID}}=4\);
the decoder has three layers, \(H=16\) heads, and head dimension \(d_h=128\),
over a vocabulary of \(|\mathcal{V}|=512\). Under a fixed beam size \(M=1024\),
this requires 533 GFLOPs per request.

\subsection{Beam Search with Dynamic Sizes}
\label{sec:beam-search-dynamic-sizes}

Beam search has two compute overheads in this workload. The first is
beam-multiplied decoder compute: each autoregressive step runs the Transformer
decoder once for every beam, so decoder FLOPs scale with the pre-configured
beam size chosen for that step. The second is
beam rearrangement: after \cd+\gtm{} filtering, beam top-\(k\)
selects the next beam's prefixes from the scored candidates, and those
selected children become the next beam array. Generated tokens, accumulated
scores, and self-attention KV state must then follow the new beam order,
making beam rearrangement a memory-operation cost in the decode loop. \grace{}
reduces these compute costs with two corresponding optimizations.

\textbf{Dynamic per-step beam sizes.} Standard beam search uses a fixed beam size
\(M\) at every step. \grace{} instead uses dynamic per-step beam sizes
\((M_1,\dots,M_{L_{\mathrm{SID}}})\). At step \(t\), the decoder runs only on \(M_t\) pre-configured
beam rows, and each row produces logits over the SID-token vocabulary
\(|\mathcal{V}|\). The \cd+\gtm{} filtering, beam score update, and top-\(k\)
selection follow the inference flow in Section~\ref{sec:inference-flow}.
The first step has \(M_1=1\) because decoding starts from a single initial
state. For intermediate steps, top-\(k\) selects the next \(\mathrm{topk}_t=M_{t+1}\) beams;
at the final step it selects the generated SIDs.

Dynamic beams avoid over-allocating scoring work near the trie root. In our
evaluation \(|\mathcal{V}|=512\), so the first step can produce at most 512 distinct prefixes.
\cd+\gtm{} filtering can reduce this further: the experiments in
Section~\ref{sec:eval-gtm-kway} measure 492.6 unique eligible beams on average at the first
position. From the second step onward,
\(M_t\cdot|\mathcal{V}|\) is much larger than \(M_{t+1}\), so top-\(k\) outputs
normally fill the next beam size. \grace{} therefore uses
\((M_1,M_2,M_3,M_4)=(1,512,1024,1024)\), rather than a fixed beam-size.
Decoder compute is approximately linear in beam size,
so this reduces model compute from 533 to 446 GFLOPs per
request in the 4-step setup, a 16.2\% reduction.

\textbf{Paged beam rearrangement.} After each top-\(k\) selection, beam rearrangement
forms the next beam state by shuffling the selected prefixes' tokens, scores,
and KV state, then appending the new token.
Generated tokens and accumulated scores are lightweight; the more demanding
data movement comes from the self-attention KV cache. A dense implementation copies
the selected beams' past prefix KV into each new beam row. At \(B\cdot M=16384\), this
dense KV copy moves about 1.6 GB and costs about 1.79 ms per step. \grace{}
instead organizes self-attention KV in a paged-attention-style block table:
each block stores one token's self-attention KV for one beam row, across all
heads. Beam rearrangement copies only block ids for previous
positions, and the next decoder step writes the current token into a fresh
block. Children that share a prefix can therefore reuse the same past block ids;
Section~\ref{sec:kernel-transformations} details the kernel layout.

\subsection{Multi-Stage Pipelined Inference}
\label{sec:pipelined-inference}

\grace{} implements the inference flow as four runtime stages that run
concurrently: (1) an input stage for request batching and host-to-GPU
transfer, (2) a feature preprocessing stage for embedding-heavy feature
preparation, including embedding bags and sequence/non-sequence preprocessing,
(3) a compute-intensive stage for the encoder's feature-interaction and
attention compute, followed by the full decoding loop with \gtm{}, and
(4) a post stage for \sid{}-to-ad lookup and exact ad-level target
filtering. Each stage runs on a Python thread with a dedicated CUDA stream.
Python GIL contention is mitigated by moving the hot path out of the Python
interpreter through TorchScript, CUDA graphs, and a C++ implementation of the
post stage.

\begin{figure}[t]
  \centering
  \includegraphics[width=0.92\columnwidth]{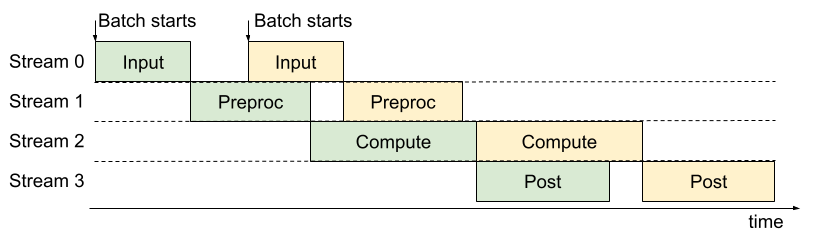}
  \caption{Multi-stage pipelined inference in \grace{}. Input, feature
  preprocessing, compute-intensive, and post stages run on separate Python
  threads and CUDA streams. The compute-intensive
  stage runs back-to-back on its own stream to keep GPU utilization high and
  hide lower-MFU stages.}
  \label{fig:pipelined-parallelism}
\end{figure}

For one batch, these stages execute in order. Across different batches, they
are staggered across streams and run in parallel, as illustrated in
Figure~\ref{fig:pipelined-parallelism}. We report Model FLOPs utilization
(MFU) as observed computational throughput divided by the maximum theoretical
compute capacity of the hardware. The feature preprocessing stage has lower
MFU, while the compute-intensive stage has high MFU and dominates compute.
Because these stages stress different parts of the hardware,
the compute-intensive stage can run back-to-back while
lower-MFU stages overlap in the same time window, improving end-to-end
GPU efficiency without changing the model computation or introducing measurable
regression from parallel execution.

\section{Generative Target Matching}

\subsection{Problem Formulation}

We distinguish catalog validity from advertiser eligibility. A valid \sid{}
exists in the catalog and maps to at least one ad; an eligible ad is one that
is matched with the user/request by the targeting rules; an eligible \sid{} is a \sid{} with at
least one eligible ad; and an eligible \sid{} prefix is a prefix whose subtree
contains at least one eligible ad.

Conventional Target Matching starts from the request attributes and returns a
set of ads by querying the targeting index, then hands the surviving ads to
the retrieval model. A generative retrieval model has no such pre-matched ad
inputs. At step \(t\), the only choice available is which next SID tokens the
beam search is allowed to explore.
\gtm{} translates the conventional target filtering contract into this
token-level control problem: a token is feasible only if the prefix it creates
still contains at least one ad that could pass target filtering for the
request.

At decode step \(t\), the current beam prefix is \(\mathbf{p}\) in the \sid{}
trie. The valid next SID tokens are the children of \(\mathbf{p}\) in this trie,
and \(v\) denotes one candidate next token. Constrained decoding admits $v$ if
the concatenated prefix \(\mathbf{p}\Vert v\), formed by appending token \(v\)
to prefix \(\mathbf{p}\), is a prefix of some valid \sid{}:
\[
\mathbf{CD}_t[v]=\mathbf{1}[v\in\mathrm{children}(\mathbf{p})].
\]
\gtm{} defines an eligibility indicator \(\mathbf{TM}_t[v]\) for each candidate
token. The request attributes used by Target Matching are denoted by
\(\mathbf{u}\), such as country, age, gender, and location.
It admits $v$ if the SID subtree rooted at $\mathbf{p}\Vert v$ contains at
least one ad eligible for request attributes $\mathbf{u}$:
\[
\mathbf{TM}_t[v]=\mathrm{match}(\mathbf{u},\mathbf{p}\Vert v).
\]
The decoder applies both constraints to the log probabilities of candidate tokens:
\[
\mathbf{L}_t[v]\leftarrow
\begin{cases}
\mathbf{L}_t[v], & \mathbf{CD}_t[v]\mathbf{TM}_t[v]=1,\\
-\infty, & \mathrm{otherwise}.
\end{cases}
\]

\(\mathbf{TM}_t[v]\) is implemented with two matcher types: bitmask matchers
for low-cardinality attributes and Bloom filter matchers for high-cardinality
attributes.

\subsection{Bitmask Matching for Low-cardinality Attributes}
\label{sec:gtm-bitmask}

For low-cardinality constraints, bitmasks encode targeting attributes such as
country, age, gender, and other user info. Each attribute has a fixed-size bit
range, where each bit position corresponds to one possible value. Let
\(\mathrm{bitmask}(a)\) encode the values allowed by ad \(a\)'s targeting rules,
and let \(\mathrm{bitmask}(\mathbf{u})\) encode the user's observed attribute
values. The full bitmask concatenates these per-attribute ranges into one packed
vector stored as a fixed-size array of 64-bit integers. We pad bitmasks to
64-bit word boundaries when needed, enabling CPU-side SIMD-vectorized bit operations.

\begin{figure}[t]
  \centering
  \includegraphics[width=\columnwidth]{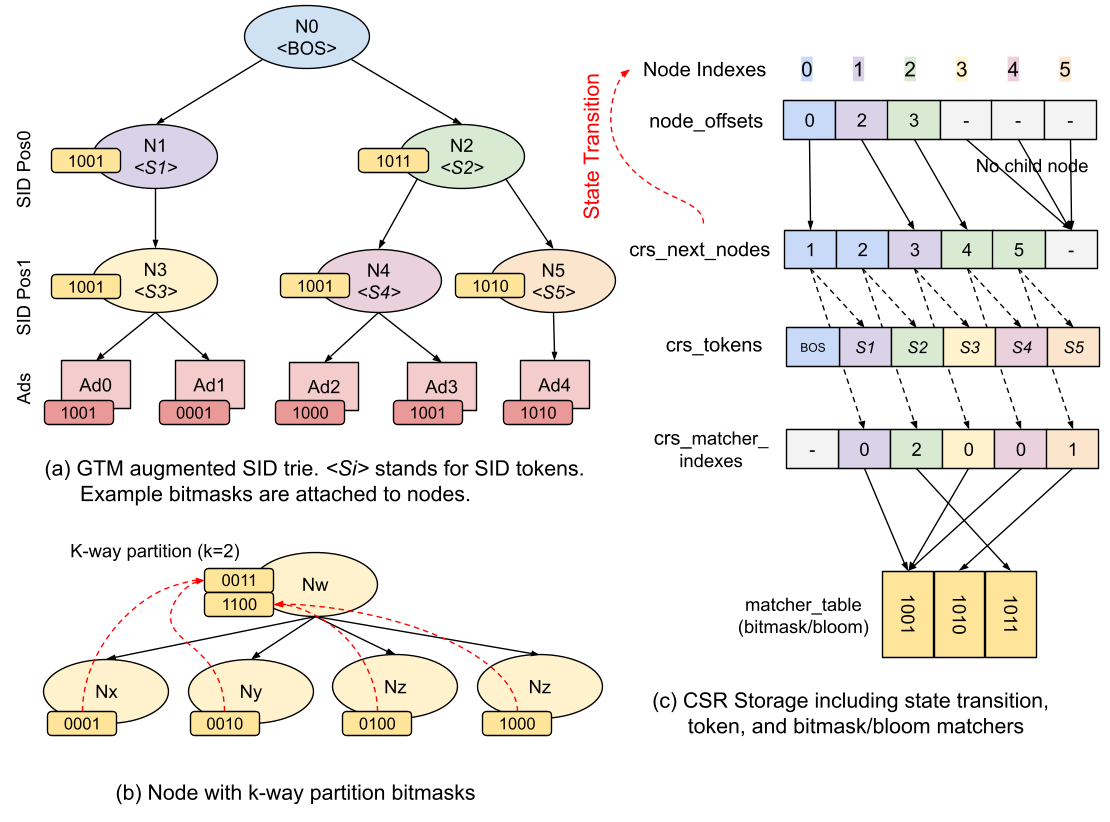}
  \caption{\gtm{} storage and k-way partitioning. Panel (a) shows subtree-union
  matcher entries on SID-trie nodes. Panel (b) applies k-way partitioning,
  splitting one broad subtree union into several partition-level unions.
  Panel (c) shows the \cd+\gtm{} index layout: CSR child-entry arrays provide
  child node ids, SID tokens, and matcher-table ids, while bitmask and Bloom
  matcher rows are stored in deduplicated tables. Appendix~\ref{app:gtm-index-layout}
  gives the concrete lookup layout.}
  \label{fig:gtm-storage}
\end{figure}

\grace{} materializes such bitmasks for SID prefixes and full SIDs.
Let \(\mathcal{A}(\mathbf{p})\)
be the ads represented by trie node \(\mathbf{p}\): for a full SID, this is
the ad cluster assigned to that SID; for an internal SID prefix, this is the
union of ads under all descendant SIDs. The stored bitmask matcher is the OR
of those ad bitmasks:
\[
\mathrm{bitmask}(\mathbf{p})=
\bigvee_{a\in\mathcal{A}(\mathbf{p})}\mathrm{bitmask}(a).
\]
Figure~\ref{fig:gtm-storage} shows the same union principle with a smaller toy
mask: an internal-node matcher is formed by OR-ing the matchers below it.
During decoding, candidate token \(v\) points to child node
\(\mathbf{p}\Vert v\), and \(\mathbf{TM}_t[v]\) evaluates the bitmask match
between \(\mathrm{bitmask}(\mathbf{u})\) and the stored node matcher
\(\mathrm{bitmask}(\mathbf{p}\Vert v)\). The packed test passes when all user
attribute bits are present in the node matcher:
\[
(\mathrm{bitmask}(\mathbf{p}\Vert v)\wedge\mathrm{bitmask}(\mathbf{u}))
=\mathrm{bitmask}(\mathbf{u}).
\]
Because the stored matcher is an OR over ads under the prefix, the test is \textbf{conservative}
and may introduce false positives: it can make a node
pass even when no single ad under that node is eligible, because different ads
may contribute different matching attribute bits.

As a small bit-level example, suppose a toy mask has two attributes:
country bits \([\mathrm{US},\mathrm{CA},\mathrm{UK}]\), age bits
\([18\text{--}24,25\text{--}34]\). If a trie node contains one ad targeting
US users in the 18--24 age bucket and another ad targeting CA users in the
25--34 age bucket, the node union matcher is
\[
\mathrm{bitmask}(\mathbf{p}) =
\underbrace{\texttt{110}}_{\mathrm{country}}\,
\underbrace{\texttt{11}}_{\mathrm{age}} .
\]
A request from a user in the US, at age 30, has
\[
\mathrm{bitmask}(\mathbf{u}) =
\underbrace{\texttt{100}}_{\mathrm{country}}\,
\underbrace{\texttt{01}}_{\mathrm{age}} ,
\]
so every attribute intersects:
\(\texttt{100}\wedge\texttt{110}\ne0\),
\(\texttt{01}\wedge\texttt{11}\ne0\). The node is kept, even though neither
ad is individually eligible: the US ad fails the age attribute, and the
25--34 ad fails the country attribute.

\textbf{k-way partition.} k-way partitioning mitigates OR-induced false
positives by splitting descendants under a trie node into \(k\) groups and
storing one union matcher per group; a request passes only if at least one
group matches all attributes. Figure~\ref{fig:gtm-storage}(b) illustrates this
partitioned layout.
We define matcher fill rate as the fraction of set bits in a stored union
matcher row. Lower fill means the union bitmask filter is less
saturated, so the matcher is more selective and less likely to keep prefixes
that are false positives of the subtree union. The tradeoff is that k-way
partitioning stores and checks more matchers for each node: a prefix is kept
if any partition matcher passes.
\begin{itemize}
  \item \textbf{Static k-way} uses a fixed \(k\) per decode position. During
  construction, each SID is assigned to the partition that gives the smallest
  post-union fill rate after adding that SID's matcher.
  \item \textbf{Dynamic k-way} chooses \(k\) separately for each trie node, i.e.,
  for each SID prefix. The builder increases \(k\) for that node until the
  union fill rate reaches a best-effort target or reaches \(k_{\max}\).
\end{itemize}
As discussed in Section~\ref{sec:inference-flow}, \gtm{} is a SID-level
filter, so generated SIDs still pass through the CPU-side exact ad-level
target matcher after SID-to-ad lookup. Section~\ref{sec:eval-gtm-kway}
therefore reports exact ad-level pass rate as the evaluation metric for
SID-level \gtm{}.

\subsection{Bloom Matching for High-cardinality Attributes}
\label{sec:gtm-bloom}

For high-cardinality constraints such as location, the value space is too large
to encode as dense bit ranges, and each ad or request may contain multiple
values. \grace{} therefore uses Bloom filter matchers. We use location as the
example: advertisers may target fine-grained geographies represented as strings,
such as city, region, or local-area strings, while a request carries user-side
location strings.

Each ad encodes its target-location set as a fixed-width \(w\)-bit Bloom filter
\(\mathrm{bloom}(a)\). In our evaluation, each Bloom matcher uses \(w=256\)
bits, stored as four 64-bit integers. A request with \(N_{\mathrm{loc}}\) user-side locations
\(\{\ell_1,\ldots,\ell_{N_{\mathrm{loc}}}\}\) encodes each location separately as
\(\{\mathrm{bloom}(\ell_1),\ldots,\mathrm{bloom}(\ell_{N_{\mathrm{loc}}})\}\).
We use the Bloom containment test
\[
\mathrm{contains}(B,\ell_j)=
\mathbf{1}\!\left[
(\mathrm{bloom}(\ell_j)\wedge B)=\mathrm{bloom}(\ell_j)
\right].
\]
A location constraint matches a request with an ad when
\(\mathrm{contains}(\mathrm{bloom}(a),\ell_j)=1\) for at least one user
location \(\ell_j\).

\grace{} materializes Bloom matchers for the same SID prefixes and full SIDs as
the bitmask matchers in Section~\ref{sec:gtm-bitmask}. Using the same \(\mathcal{A}(\mathbf{p})\) notation,
the stored Bloom matcher for node \(\mathbf{p}\) is the OR of the Bloom filters
for ads represented by that node:
\[
\mathrm{bloom}(\mathbf{p})=
\bigvee_{a\in\mathcal{A}(\mathbf{p})}\mathrm{bloom}(a).
\]
During decoding, \(\mathbf{TM}_t[v]\) uses the same child-node lookup pattern
as bitmask matching, but the Bloom predicate is ANY-of-many: it keeps
\(\mathbf{p}\Vert v\) as soon as
\(\mathrm{contains}(\mathrm{bloom}(\mathbf{p}\Vert v),\ell_j)=1\) for any
request location \(\ell_j\).
As in Section~\ref{sec:gtm-bitmask}, the subtree OR is conservative and may introduce
false positives; Bloom hash collisions add another source of false positives.

The same k-way partitioning from Section~\ref{sec:gtm-bitmask} applies to Bloom matchers: instead
of storing one broad OR-ed Bloom filter for a prefix, the index can store
several partition-level Bloom filters and keep the prefix if any partition
matches a request location. This can reduce false positives from subtree
unions, at the cost of more matcher entries and decode-time checks.
Section~\ref{sec:eval-gtm-kway} evaluates this tradeoff. Because SID-level
Bloom filtering remains conservative, \grace{} still runs exact ad-level
Bloom filtering to remove ineligible ads.

\subsection{Storage Layout and Lookup}
\label{sec:gtm-index}

Following the constrained decoding index introduced by STATIC~\cite{static},
\grace{} stores valid \sid{} sequences as a prefix trie.
Appendix~\ref{app:gtm-index-layout} gives the concrete storage layout. During
decoding, each active beam corresponds to one trie prefix/node and carries the
integer node id for that prefix in addition to its decoded tokens and
accumulated score.

\grace{} stores trie children in compressed sparse row (CSR) form. The
\emph{node offsets} array maps each trie node to a contiguous child-entry range;
for node \(r\), the range between consecutive offsets identifies all valid
next-token children. For every child-entry id \(j\) in this range, the
\emph{CSR tokens} array gives the candidate SID token, the
\emph{CSR child nodes} array gives the child node id to carry if the candidate
survives, and the \emph{bitmask index} and \emph{Bloom index} arrays identify
the matcher rows used for eligibility checking.
Figure~\ref{fig:gtm-storage}(c) visualizes this \cd+\gtm{} storage layout, where
CSR child-entry arrays provide constrained decoding transitions and matcher ids
point to deduplicated bitmask/Bloom tables for eligibility matching.

At step \(t\), after the decoder produces log probabilities over the SID
vocabulary, the \cd+\gtm{} kernel enumerates valid children for each active beam.
It then evaluates the child matchers against request-side user attributes
\(\mathbf{u}\), encoded as bitmask and Bloom filter rows. The resulting mask
removes invalid or ineligible tokens before beam top-\(k\), and the selected
child node ids become the node ids for the next decode step.
Section~\ref{sec:gtm-kernel} describes the GPU
implementation of this lookup and matching path.

\textbf{Deduplicated bitmask/Bloom storage.} In practice, many child entries
share identical bitmask or Bloom matcher rows. \grace{} therefore deduplicates matcher storage:
the matcher-table ids read from the CSR range point into shared tables of unique
bitmask and Bloom matcher rows. Table~\ref{tab:index-summary} reports the
index distribution and matcher storage before and after deduplication.
The largest savings occur at p3, the full-SID
transition level: bitmask storage drops from 1645 MB to 228 MB, and Bloom
storage drops from 940 MB to 162 MB.

The k-way partitioning layout uses the same deduplicated-table format, but a
trie node can store one or more matcher ids. In static k-way, the number of
partitions is fixed by decode position, so matcher ids are stored in a
fixed-width per-node layout. In dynamic k-way, each node can choose a different number of
partitions, so matcher ids are stored in a CSR-style jagged format with a flat
matcher id array and per-row offsets.

\begin{table}[t]
  \caption{Index summary and GTM matcher storage by decode position. Here p1/p2/p3 denote non-root trie transition levels after prefixes of length 1/2/3; for \(L_{\mathrm{SID}}=4\), p3 reaches full SIDs. GTM matcher storage values are before/after deduplication in MB.}
  \label{tab:index-summary}
  \centering
  \widetable
  \begin{tabular*}{\columnwidth}{@{\extracolsep{\fill}}ll@{}}
    \toprule
    Metric & Quantity \\
    \midrule
    SIDs & 30M \\
    Ads/SID p50/p90/p99 & 1 / 2 / 11 \\
    \midrule
    Bitmask width & 7 int64 values \\
    Bloom width & 4 int64 values \\
    \midrule
    Bitmask storage p1/p2/p3 & 11/10, 620/175, 1645/228 \\
    Bloom storage p1/p2/p3 & 6/5, 354/112, 940/162 \\
    \bottomrule
  \end{tabular*}
\end{table}

\section{Decoder Kernel Design}
\label{sec:kernel-transformations}

Using the encoder-decoder setup from Section~\ref{sec:encoder-decoder-model}, the decoder kernel benchmark
runs with batch size \(B=16\). Under the fixed beam size \(M=1024\) schedule,
the first step has \(B\) active rows; later steps have \(B\cdot M=16384\) rows.
This wide-beam, short-sequence shape is poorly
matched to general SDPA kernels: cross-attention has many one-token beam
queries that share request KV, while self-attention has one-token queries over
only the partial \sid{} prefix. This differs from common LLM serving regimes:
prefill typically processes long query and key sequences with high compute
intensity, while decode uses one-token queries over much longer KV caches and
is often dominated by memory bandwidth.

In \grace{}, the pressure comes from thousands of small attention operations,
memory-coalescing behavior, and beam rearrangement rather than from long
sequence attention. The following subsections describe the corresponding decoder
optimizations: a cross-attention layout that lets beams share request-level KV,
short-sequence self-attention kernels, and paged self-attention KV for beam
rearrangement.

\grace{} also reduces framework overhead around these kernels. We use
\textnormal{torch.compile} max-autotune to let the compiler auto-fuse surrounding
PyTorch operations where possible, while Section~\ref{sec:self-attn-kernel} describes the manual
fusion inside the custom self-attention kernel. The full fixed-length SID
decoding loop is then captured as a CUDA graph, including decoder forward,
\cd+\gtm{}, top-\(k\), and beam rearrangement.

\subsection{Cross-Attention Beam-as-Query Layout}
\label{sec:cross-attn-layout}

Before the cross-attention reshape, a general SDPA layout would use
\[
Q\in\mathbb{R}^{B M\times H\times 1\times d_h},\qquad
K,V\in\mathbb{R}^{B M\times H\times T_{\mathrm{ctx}}\times d_h}.
\]
Here \(BM\) denotes batch size \(B\) times beam size \(M\). This treats the
decode as \(BM\) independent single-query attention operations.
It misses the beam search structure: the \(M\) beams for one request have
different queries but attend to the same user-context KV. \grace{} keeps
cross-attention \(K,V\) in their natural request-level layout,
\(\mathbb{R}^{B\times H\times T_{\mathrm{ctx}}\times d_h}\), and permutes
\[
Q:\mathbb{R}^{B M\times H\times 1\times d_h}
\rightarrow
\mathbb{R}^{B\times H\times M\times d_h},
\]
so the beams of one request become the query-sequence dimension. The attention
call is then \(M\) beam queries over one shared user context, rather than \(M\)
unrelated one-query calls. The reshape keeps one copy of the request KV,
creates a better-shaped attention operation, and reuses the same user-context KV
loads across all beams for that request.

At decode step \(t\), with \(T_q=1\),
the beam-as-query reshape is:
\[
\begin{array}{l}
q=\mathrm{view}(Q,\;[B,M_t,H,T_q,d_h]),\\
q=\mathrm{permute}(q,\;[0,2,1,3,4])
  \rightarrow[B,H,M_t,T_q,d_h],\\
q=\mathrm{reshape}(q,\;[B,H,M_tT_q,d_h]),\\
x=\mathrm{SDPA}(q,K,V)
  \rightarrow[B,H,M_tT_q,d_h],\\
x=\mathrm{reshape}(x,\;[B,H,M_t,T_q,d_h]),\\
x=\mathrm{permute}(x,\;[0,2,1,3,4])
  \rightarrow[B,M_t,H,T_q,d_h],\\
X=\mathrm{reshape}(x,\;[BM_t,H,T_q,d_h]).
\end{array}
\]
Thus SDPA sees \(M_tT_q\) query positions for each request, while the output
\(X\) restores the original beam-major layout. Appendix~\ref{app:cross-attn-correctness}
gives the correctness argument for this layout transformation.

\subsection{Coalesced Short-Sequence Self-Attention}
\label{sec:self-attn-kernel}

Self-attention also has single-token queries, but its
\(K,V\in\mathbb{R}^{BM\times H\times T_{\mathrm{self}}\times d_h}\) are
beam-specific because each beam has a different generated prefix. Therefore
the cross-attention KV-broadcast trick does not apply. Instead, the exploitable
structure is that many flattened batch/beam rows have the same short sequence
length. \grace{} uses a Triton SDPA kernel tailored to this decode shape: it
coalesces multiple beam rows into one kernel tile, computes a larger attention
tile, and applies a block-diagonal mask so each row attends only to its own
cached prefix.

For example, suppose a tile coalesces \(G=4\) beam rows with
\(T_{\mathrm{self}}=3\). This represents four independent \(1\times 3\)
attentions, but the Triton kernel evaluates one \(4\times 12\) tile and masks
cross-row entries. In implementation, \(G\) is picked by Triton autotuning from
a small set of candidate coalescing factors.
\[
\resizebox{\columnwidth}{!}{$
\begin{array}{c|ccc|ccc|ccc|ccc}
 & k_{0,0} & k_{0,1} & k_{0,2}
 & k_{1,0} & k_{1,1} & k_{1,2}
 & k_{2,0} & k_{2,1} & k_{2,2}
 & k_{3,0} & k_{3,1} & k_{3,2} \\
\midrule
q_0 & 1 & 1 & 1 & 0 & 0 & 0 & 0 & 0 & 0 & 0 & 0 & 0 \\
q_1 & 0 & 0 & 0 & 1 & 1 & 1 & 0 & 0 & 0 & 0 & 0 & 0 \\
q_2 & 0 & 0 & 0 & 0 & 0 & 0 & 1 & 1 & 1 & 0 & 0 & 0 \\
q_3 & 0 & 0 & 0 & 0 & 0 & 0 & 0 & 0 & 0 & 1 & 1 & 1
\end{array}
$}
\]
Here \(q_i\) is the current query token for beam \(i\), and
\(k_{i,j}\) is self-attention key \(j\) for the same beam. Entries marked
0 are set to \(-\infty\) before softmax. This mask is not a causal mask:
causality is already enforced by exposing only the beam prefix
\(0,\ldots,t\). The mask only prevents different beam rows from attending to
one another. This transformation spends extra arithmetic: the useful fraction is \(1/G\).
This is still a net win because
\(T_{\mathrm{self}}\) is only a few \sid{} tokens, while the baseline executes
many tiny \(1\times T_{\mathrm{self}}\) attentions with poor occupancy, small
memory transactions, and weak tensor-core utilization. Coalescing turns them
into fewer, larger, regular tiles with more coalesced memory access.

\textbf{Kernel fusion.}
At higher optimization levels, the decoder further fuses the non-GEMM middle of
decoder self-attention. The QKV projection and output projection remain cuBLAS
GEMMs; the fused Triton kernel consumes the QKV projection output directly,
writes the current token's \(k,v\) into the cache, gathers the short cached
prefix, computes the masked block-diagonal attention, and returns the output.
This avoids separate cache-write, reshape, SDPA, and context materialization
steps.

\subsection{Paged Self-Attention KV Cache}
\label{sec:paged-kv-cache}

Section~\ref{sec:beam-search-dynamic-sizes} introduces the paged beam
rearrangement idea. Here we specify the self-attention KV layout used by the
kernel. Following PagedAttention~\cite{pagedattention}, \grace{} stores KV
blocks in a flat pool and uses a block table
\(P\in\mathbb{Z}^{BM_t\times L_{\mathrm{SID}}}\) to map logical beam rows and
SID positions to physical KV blocks. A block stores the self-attention KV for
one generated token of one beam row. Thus \(P_{i,t}\) maps logical beam row
\(i\) and SID position \(t\) to the physical KV block for that token.

Let \(p_i\) be the parent row selected by beam \(\mathrm{topk}_t\) for new beam
row \(i\). A dense implementation copies the parent's cached KV bytes into
row \(i\). \grace{} instead copies only the parent's block ids for the existing
prefix, \(P^{\mathrm{new}}_{i,j}=P_{p_i,j}\) for \(0\le j<t\), and writes the current
token's KV into a fresh block. During attention, the kernel follows the block
ids in prefix order to gather the physical KV blocks. This makes KV reads less
contiguous than a dense reordered cache, but replaces large KV-cache copies
with small integer-table copies; beams that share a parent also share the same
past block ids. Section~\ref{sec:kernel-performance} and Table~\ref{tab:self-attn-microbench} report the
latency impact, and Table~\ref{tab:decoder-opt-levels} shows the net
end-to-end gain.

\subsection{GTM Kernel}
\label{sec:gtm-kernel}

The \cd+\gtm{} kernel implements the decode-time lookup path from
Section~\ref{sec:gtm-index} and Appendix~\ref{app:gtm-index-layout}.
We implement this path as CUDA kernels rather than Triton kernels to
get finer control over shared-memory staging, per-candidate thread assignment,
and early exit behavior. Each CUDA block handles one beam, and threads within
the block process the beam's child-entry range in parallel. Since all beams
expanded from the same user request share the same user-side targeting
attributes, the kernel maps the beam row back to the corresponding row in the
batched user-attribute tensors and stages those masks in shared memory once.
All next-token candidates for that beam then reuse the staged user-attribute
masks instead of reloading them.

The kernel evaluates the predicates from Sections~\ref{sec:gtm-bitmask}
and~\ref{sec:gtm-bloom} in an order that
reduces work. It checks the bitmask matcher first, comparing packed int64 values
and exiting on the first failed int64 comparison; Bloom matching runs only for
candidates that pass the bitmask check. For Bloom matching, the kernel scans
the cached request Bloom rows, one row per user-side location \(\ell_j\). For
each location, it compares the request-location Bloom int64 values against the
candidate Bloom matcher and stops the inner scan as soon as a required int64
value is not contained. The outer scan stops as soon as any user location is
contained in the candidate matcher, because one matched location is sufficient
for the ANY-of-many location predicate.

The early exits reduce instruction count and matcher memory traffic, although a
CUDA block's completion time is still bounded by the slowest active candidate
processed by its threads.

\section{Evaluation}
\label{sec:evaluation}

\subsection{Setup}

We evaluate \grace{} by running the inference flow in
Section~\ref{sec:inference-flow}. The evaluation has two parts. First, we
measure exact ad-level target matching pass rate after generated SIDs are
expanded and filtered by the CPU-side ad-level matcher; this evaluates the
effectiveness of SID-level \gtm{} under the final ad-level eligibility check.
Second, we measure serving performance with kernel microbenchmarks and
end-to-end latency for the full inference path.

The filtering study uses the model evaluation dataset for user features,
synthetic user-location data with 64 random locations per user, and the
30M-\sid{} index summarized in Table~\ref{tab:index-summary}. Performance
measurements use the model configuration from Section~\ref{sec:encoder-decoder-model} and run on an
NVIDIA GH200 Grace Hopper Superchip~\cite{gh200}.

\subsection{Ad-Level Target Matching Pass Rate}
\label{sec:eval-gtm-kway}

This subsection compares four decode-time filtering modes. \textbf{CD only}
uses constrained decoding for catalog-valid SIDs but disables SID-level
targeting. \textbf{CD+\gtm{}} enables bitmask and Bloom matchers without k-way
partitioning, so each trie entry has one union matcher for each matcher type.
The matcher rows are stored in the deduplicated format described in
Section~\ref{sec:gtm-index}. \textbf{Static k-way} and \textbf{Dynamic k-way} use the
partitioning variants from Section~\ref{sec:gtm-bitmask} for bitmask matchers and Section~\ref{sec:gtm-bloom}
for Bloom matchers. For static k-way, both bitmask and Bloom use
\(k=(64,64,32,8)\) across the four decode positions.
For dynamic k-way, each trie node chooses \(k\) independently
with \(k_{\max}=32\) for both bitmask and Bloom, using a best-effort target
fill rate of 0.25 defined in Section~\ref{sec:gtm-bitmask},

Table~\ref{tab:kway-fill} reports the resulting GTM matcher fill rates.
Root is the start-token transition from the empty prefix, and Pos1--Pos3 are transitions after SID
prefixes of length 1, 2, and 3. Since \(L_{\mathrm{SID}}=4\), Pos3 produces complete SIDs.
Static k-way substantially reduces matcher saturation for both
bitmask and Bloom matchers, while dynamic k-way is not effective
because it often selects small \(k\).

\begin{table}[t]
  \caption{Mean GTM matcher fill rate by decode position. Fill rate is the
  fraction of 1 bits in the bitmask or Bloom matcher; lower is more selective.}
  \label{tab:kway-fill}
  \centering
  \widetable
  \begin{tabular*}{\columnwidth}{@{\extracolsep{\fill}}llrrrr@{}}
    \toprule
    Mask & Method & Root & Pos1 & Pos2 & Pos3 \\
    \midrule
    Bitmask & CD+\gtm{} & 0.734 & 0.283 & 0.139 & 0.128 \\
    Bitmask & Static k-way & \textbf{0.563} & \textbf{0.097} & \textbf{0.010} & \textbf{0.016} \\
    Bitmask & Dynamic k-way & 0.627 & 0.283 & 0.142 & 0.128 \\
    \midrule
    Bloom & CD+\gtm{} & 0.979 & 0.475 & 0.069 & 0.048 \\
    Bloom & Static k-way & \textbf{0.933} & \textbf{0.070} & \textbf{0.004} & \textbf{0.006} \\
    Bloom & Dynamic k-way & 0.976 & 0.333 & 0.080 & 0.048 \\
    \bottomrule
  \end{tabular*}
\end{table}

Table~\ref{tab:kway-pass} reports the exact ad-level pass rate after generated
SIDs are mapped to ads and filtered by the CPU-side ad-level target matcher.
To analyze where the pass-rate changes come from, we bucket requests by the
number of generated ads after SID-to-ad lookup and before ad-level filtering.
These buckets measure generated-ad volume before exact filtering and show how
\gtm{} changes the generated-ad distribution.
CD+\gtm{} shifts requests from the \(<5\)k bucket to larger
buckets: the \(<5\)k share drops from 50.78\% to 31.84\%, while the 10k+ share
rises from 12.50\% to 29.10\%. This means decode-time targeting produces more
ads before the exact ad-level filtering stage, because \gtm{} selects different
SIDs than CD-only decoding.

A higher final pass rate means that \gtm{} more effectively focuses decoding on eligible SIDs and generates
fewer ineligible ads.  Comparing CD only with CD+\gtm{} shows the main effectiveness result.
The all-request final pass rises from 23.55\% to 40.42\%. The gain is largest in
the lower- and mid-volume buckets: \(<5\)k improves from 20.83\% to 56.33\%,
and 5k--9.9k improves from 21.92\% to 41.80\%. Both matcher types contribute:
for \(<5\)k, bitmask pass improves from 57.72\% to 76.82\%, and Bloom pass
after bitmask improves from 36.09\% to 73.33\%.

\begin{table}[t]
  \caption{Bucketized ad-level pass-rate comparison. Buckets are generated ads per request after SID lookup and before ad-level filtering; Users is the request fraction in each bucket.}
  \label{tab:kway-pass}
  \centering
  \widetable
  \begin{tabular*}{\columnwidth}{@{\extracolsep{\fill}}lrrrr@{}}
    \toprule
    Method & Users & Bitmask & Bloom & Final \\
    \midrule
    \multicolumn{5}{@{}l}{\textit{\(<5\)k generated ads/user}} \\
    CD only & 50.78\% & 57.72\% & 36.09\% & 20.83\% \\
    CD+\gtm{} & 31.84\% & \textbf{76.82\%} & 73.33\% & \textbf{56.33\%} \\
    Static k-way & 31.84\% & 76.34\% & 72.90\% & 55.65\% \\
    Dynamic k-way & 31.84\% & 76.09\% & \textbf{73.44\%} & 55.88\% \\
    \midrule
    \multicolumn{5}{@{}l}{\textit{5k--9.9k generated ads/user}} \\
    CD only & 36.72\% & 54.91\% & 39.92\% & 21.92\% \\
    CD+\gtm{} & 39.06\% & 64.14\% & 65.17\% & 41.80\% \\
    Static k-way & 38.09\% & \textbf{64.98\%} & \textbf{65.63\%} & \textbf{42.65\%} \\
    Dynamic k-way & 38.87\% & 64.58\% & 64.88\% & 41.90\% \\
    \midrule
    \multicolumn{5}{@{}l}{\textit{10k+ generated ads/user}} \\
    CD only & 12.50\% & \textbf{62.28\%} & 45.95\% & 28.62\% \\
    CD+\gtm{} & 29.10\% & 56.02\% & 63.46\% & 35.55\% \\
    Static k-way & 30.08\% & 55.44\% & 63.29\% & 35.09\% \\
    Dynamic k-way & 29.30\% & 56.12\% & \textbf{63.67\%} & \textbf{35.73\%} \\
    \midrule
    \multicolumn{5}{@{}l}{\textit{All requests}} \\
    CD only & 100.00\% & 57.81\% & 40.74\% & 23.55\% \\
    CD+\gtm{} & 100.00\% & 61.52\% & \textbf{65.70\%} & 40.42\% \\
    Static k-way & 100.00\% & 61.31\% & 65.68\% & 40.27\% \\
    Dynamic k-way & 100.00\% & \textbf{61.58\%} & 65.69\% & \textbf{40.45\%} \\
    \bottomrule
  \end{tabular*}
\end{table}

Overall, k-way partitioning does not materially improve final pass rate over
unpartitioned CD+\gtm{}. Although partitioning reduces matcher fill and makes each
partition matcher more selective, decoding keeps a prefix if any partition
matches, so the results from multiple matchers are unioned.
Considering the extra storage and kernel overhead of checking multiple
partitions, the unpartitioned CD+\gtm{} layout is the preferred operating point.

\subsection{Kernel-Level Performance}
\label{sec:kernel-performance}

After \gtm{} improves candidate quality, the system still has to fit a tight
P99 \(<100\) ms end-to-end inference latency budget. About 30 ms is used to
accumulate batches of 16 users, leaving roughly 70 ms for the four runtime
stages described in Section~\ref{sec:pipelined-inference}: input, feature
preprocessing, compute-intensive inference, and post-processing.

We first measure \grace{}'s attention-kernel improvements because attention was
the bottleneck and provides the largest latency gains. The cross-attention
beam-as-query layout
from Section~\ref{sec:cross-attn-layout} reduces FlashAttention-2 (FA2) latency from 9.6 ms to 0.097 ms
(98.5\(\times\)) and FlashAttention-3 (FA3) latency from 6.9 ms to 0.101 ms
(68.0\(\times\)), as shown in Table~\ref{tab:cross-attn-q-shape}.

For self-attention, Table~\ref{tab:self-attn-microbench} reports three
variants that correspond to the optimizations in Section~\ref{sec:kernel-transformations}. \textbf{Coalesced}
is the short-sequence Triton kernel from Section~\ref{sec:self-attn-kernel}, which groups beam rows
for coalesced block-diagonal attention. \textbf{Fused} adds the decode-time
self-attention fusion from Section~\ref{sec:self-attn-kernel} while using dense KV cache.
\textbf{Paged} adds the block table KV layout from Section~\ref{sec:paged-kv-cache}, so beam
rearrangement avoids copying past KV. These variants reduce per-step latency from 5.5--6.0 ms
with FA2/FA3 to 0.216--0.238 ms with the Paged kernel.
NVIDIA Nsight Compute (NCU) counters show the
mechanism: at \(t=3\), FA2 executes 2.48B instructions and reaches only
2.38\% HBM bandwidth, while Paged executes 43.8M
instructions and reaches 84.46\% HBM bandwidth.

\begin{table}[t]
  \caption{Cross-attention Q-reshape benchmark on GH200. Latencies are in ms. \(R=16{\times}1024\).}
  \label{tab:cross-attn-q-shape}
  \centering
  \widetable
  \begin{tabular*}{\columnwidth}{@{\extracolsep{\fill}}llllr@{}}
    \toprule
    Layout & $Q$ & $K,V$ & FA2 & FA3 \\
    \midrule
    Before & $[R,16,1,128]$ & $[R,16,128,128]$ & 9.6 & 6.9 \\
    After & $[16,16,1024,128]$ & $[16,16,128,128]$ & 0.097 & 0.101 \\
    Speedup & -- & -- & 98.5$\times$ & 68.0$\times$ \\
    \bottomrule
  \end{tabular*}
\end{table}

\begin{table}[t]
  \caption{Self-attention GH200 microbenchmark. Latencies are P50 ms. Coalesced, Fused, and Paged cells show latency with speedup over FA2 in parentheses (FA3 is slower than FA2 in this shape). The lower block reports representative NCU counters at \(t=3\); HBM (\%) is achieved HBM bandwidth as a percentage of peak.}
  \label{tab:self-attn-microbench}
  \centering
  \scriptsize
  \setlength{\tabcolsep}{0.8pt}
  \renewcommand{\arraystretch}{1.0}
  \begin{tabular*}{\columnwidth}{@{\extracolsep{\fill}}crrrrr@{}}
    \toprule
    \(t\) & FA2 & FA3 & Coalesced & Fused & Paged \\
    \midrule
    0 & 5.563 & 5.998 & 0.474 (11.7$\times$) & 0.222 (25.1$\times$) & 0.216 (25.8$\times$) \\
    1 & 5.565 & 6.024 & 0.480 (11.6$\times$) & 0.223 (25.0$\times$) & 0.224 (24.9$\times$) \\
    2 & 5.578 & 5.986 & 0.488 (11.4$\times$) & 0.235 (23.8$\times$) & 0.233 (24.0$\times$) \\
    3 & 5.566 & 5.983 & 0.501 (11.1$\times$) & 0.269 (20.7$\times$) & 0.238 (23.4$\times$) \\
    \midrule
    TFLOP/s & 0.096 & 0.090 & 1.072 & 1.994 & 2.254 \\
    HBM (\%) & 2.38 & 2.22 & 29.36 & 75.52 & 84.46 \\
    Instr. & 2.48B & 2.13B & 57.7M & 35.8M & 43.8M \\
    Occ. (\%) & 12.19 & 14.06 & 91.24 & 36.67 & 36.44 \\
    \bottomrule
  \end{tabular*}
\end{table}

Table~\ref{tab:decoder-opt-levels} reports both full-model latency and
decoder-only latency. Full-model latency uses the end-to-end inference path
described above, while the decoder column isolates the decode loop. The first
block reports the fixed-beam optimization sweep:
cross-attention reshaping is the largest single step, reducing decoder P50 from
196.7 ms to 69.3 ms;
self-attention kernel optimization brings it to 20.3 ms; paged KV cache
reaches 17.8 ms.

The second block switches from the fixed-beam L4 setup to the more efficient
dynamic-beam schedule \([1,512,1024,1024]\). D0 reduces decoder latency to
15.5 ms. D1--D3 then add lookup, \cd{}, and \gtm{}
kernels for SID bitmask and Bloom matching. The row with both bitmask and
Bloom \gtm{} remains within the end-to-end inference latency budget: D3
full-model P99 is 53.6 ms, leaving room inside the 70 ms compute
window after batch accumulation.

We report decoder MFU, computed from per-request model FLOPs, batch size, and
batch mean decoder latency: \(\mathrm{MFU} =
\mathrm{FLOPs}_{req} \cdot B / (t_{mean}\cdot \mathrm{peak})\). Using the
dynamic-beam FLOP count in Table~\ref{tab:decoder-opt-levels}
\((0.446\) TFLOPs/request) and the GH200
Hopper GPU BF16 peak of 989 TFLOP/s, D0 corresponds to 46.6\% decoder MFU.
The row with \gtm{} matching performs the same model FLOPs, but adds non-FLOPs
\gtm{} matching work inside the decoder loop, reducing decoder MFU to 28.8\%.
With the pipelined inference design in Section~\ref{sec:pipelined-inference},
this compute-intensive decoder stage runs back to back on its own stream and can
hide lower-MFU stages while dominating overall accelerator utilization.

\begin{table}[t]
  \caption{Decoder E2E latency with batch size 16. Full and Decoder columns report P50/P99 in ms; Full includes all runtime stages in Section~\ref{sec:pipelined-inference}, and Decoder isolates the decode loop. The first block is a fixed-beam optimization sweep; the second block uses dynamic beam sizes \((1,512,1024,1024)\) and adds lookup and \gtm{} matching.}
  \label{tab:decoder-opt-levels}
  \centering
  \footnotesize
  \setlength{\tabcolsep}{2.0pt}
  \begin{tabular*}{\columnwidth}{@{\extracolsep{\fill}}llrr@{}}
    \toprule
    Case & Increment & Full & Decoder \\
    \midrule
    \multicolumn{4}{@{}l}{\textit{\(M_t=[1,1024,1024,1024]\)}} \\
    \multicolumn{4}{@{}l}{FLOPs/request: 0.533 TFLOPs} \\
    L0 & Baseline FA2 & 214.8/220.1 & 196.7/197.7 \\
    L1 & + cross-attn reshape & 86.5/87.9 & 69.3/69.4 \\
    L2 & + self-attn opt. & 38.5/40.8 & 20.3/20.4 \\
    L3 & + self-attn fusion & 39.3/45.2 & 20.4/20.5 \\
    L4 & + paged KV cache & 34.3/40.5 & 17.8/17.8 \\
    \midrule
    \multicolumn{4}{@{}l}{\textit{\(M_t=[1,512,1024,1024]\)}} \\
    \multicolumn{4}{@{}l}{FLOPs/request: 0.446 TFLOPs} \\
    D0 & L4 with dynamic beams & 34.9/35.8 & 15.5/15.8 \\
    D1 & + CD + SID to Ads Lookup & 38.9/40.9 & 16.4/16.5 \\
    D2 & + GTM bitmask & 41.4/45.0 & 16.6/16.8 \\
    D3 & + GTM bitmask + Bloom & 51.0/53.6 & 25.0/27.3 \\
    \bottomrule
  \end{tabular*}
\end{table}

\section{Discussion and Future Work}

Future work will improve \gtm{} by adding more categories of bitmask and Bloom
filter constraints and additional constraint types for more complex
target matching rules, while keeping the compatibility
with kernel optimizations. Kernel and model efficiency work includes decoder
scaling, Mixture-of-Experts (MoE), low-level kernel optimizations using
TLX~\cite{tlx}, model compression including quantization, pruning, and
sparsity, and automated kernel authoring~\cite{kernelevolve}.

A longer-term direction is extending \grace{} to optimize compute for
LLM-based generative recommenders in real-time ads recommendation.
This requires combining eligibility-aware decoding with efficient decoding
techniques, prompt engineering, model compression, and scheduling improvements
so language models can meet ads retrieval latency and compute-cost
requirements.

\section{Conclusion}

\grace{} addresses the two serving challenges created by generative ads
retrieval. For eligibility, \gtm{} moves target matching into the
autoregressive loop by combining catalog-valid constrained decoding with
personalized masks over SID prefixes. This focuses beam search on eligible SIDs
during decoding, so generated ads pass the downstream exact ad-level target
matcher at a much higher rate. For compute, \grace{} targets lightweight
encoder-decoder Transformers and redesigns the wide-beam, short-sequence
decoder with custom kernels and beam search optimizations for cross-attention,
self-attention, KV cache, and decoding beam state. Together, these changes
improve ads target filtering and keep ads generative retrieval within
latency and compute requirements.

\appendix

\section*{Appendix}

\section{GTM Index Storage and Lookup}
\label{app:gtm-index-layout}

This appendix gives the concrete storage layout used by the \cd+\gtm{} index,
including constrained decoding state transitions and eligibility matchers. For
each trie node, all valid next-token children are stored in one contiguous range
of the index arrays.

\begingroup
\scriptsize
\begin{verbatim}
GTM SID Trie
  vocab_size: int
  sid_length: int
  num_nodes: int
    # Number of integer trie node ids.

  node_offsets: int32[num_nodes + 1]
    # Children of node i:
    # [node_offsets[i], node_offsets[i + 1])

  csr_child_nodes: int32[num_child_entries]
    # Child node id for each child entry.

  csr_tokens: int16[num_child_entries]
    # SID token that reaches the child node.

  bitmask_index: int32[num_child_entries]
  bloom_index: int32[num_child_entries]
    # Matcher-table row ids.

  bitmask_table: uint64[num_bitmask_rows][W_bm]
  bloom_table: uint64[num_bloom_rows][W_bl]
    # Packed, deduplicated matcher rows.
\end{verbatim}
\endgroup

For a current trie node, lookup scans one contiguous child-entry range:

\begingroup
\scriptsize
\begin{verbatim}
Lookup(node_id, user_bitmask, user_blooms):
  start = node_offsets[node_id]
  end   = node_offsets[node_id + 1]

  for j in [start, end):
    child_node = csr_child_nodes[j]
    token      = csr_tokens[j]

    bm = bitmask_table[bitmask_index[j]]
    bl = bloom_table[bloom_index[j]]

    bitmask_ok = match(user_bitmask, bm)
    bloom_ok   = any(match(q, bl) for q in user_blooms)

    if bitmask_ok and bloom_ok:
      emit(token, child_node)
\end{verbatim}
\endgroup

At runtime, the GPU parallelizes the lookup and match across active beams and child
entries.

\section{Cross-Attention Layout Correctness}
\label{app:cross-attn-correctness}

The beam-as-query reshape in Section~\ref{sec:cross-attn-layout} is a
layout-only transformation. Assume flattened beam rows are request-major, so
row \(r=bM_t+m\) corresponds to request \(b\) and beam \(m\). The forward
reshape maps
\[
Q_{bM_t+m,h,\tau,:}\mapsto q_{b,h,mT_q+\tau,:}.
\]
Thus, for each request \(b\), SDPA sees the \(M_tT_q\) beam-query positions as
one query sequence and attends them to the same request-level
\(K_{b,h,:,:}\) and \(V_{b,h,:,:}\). The inverse reshape maps
\[
x_{b,h,mT_q+\tau,:}\mapsto X_{bM_t+m,h,\tau,:},
\]
which restores the original flattened beam-row order. Therefore the
transformation changes only the tensor layout seen by SDPA; it does not mix
requests or reorder the final beam outputs.

\bibliographystyle{mlsys2025}
\bibliography{references}

\end{document}